\documentclass[11pt]{amsart}
\usepackage{geometry}                
\usepackage{graphicx}
\usepackage{amssymb}
\usepackage{epstopdf}
\newcommand{\one}{\mathbf{1}}
\newcommand{\pseudo}[1]{#1^{\dagger}}
\title[High dimensional random walks can appear low dimensional]{High dimensional random walks can appear low dimensional: application to influenza H3N2 evolution}
\author{James Moore}
\author{Hasan Ahmed}

\begin{document}
\maketitle

\begin{abstract}
One important feature of the mammalian immune system is the highly specific binding of antigens to antibodies.
Antibodies generated in response to one infection may also provide some level of cross immunity to other infections. One model to describe this cross immunity is the notion of antigenic space, which assigns each antibody and each virus a point in $\mathbb{R}^n$. Past studies
have suggested the dimensionality of antigenic space, $n$, may be small.
In this study we show that data from hemagglutination assays suggest a high dimensional random walk (or self avoiding random random walk).
The discrepancy between our result and prior studies is due to the fact that random walks can appear low dimensional according to a variety of analyses.
including principal component analysis (PCA)
and multidimensional scaling (MDS).
\end{abstract}

\vskip 1cm
\section{Introduction}
\subsection{Antigenic Space}
During a viral infection, antibodies bind to viral antigens by recognizing specific epitopes on their surface. The same antibody may provide protection against
other strains of the same virus if the antigens are not too dissimilar. In 1979, Alan Perelson and George Oster defined the idea of antigenic space \cite{perelson1979theoretical}.
They supposed that each antibody and antigen might be described by a vector in $\mathbb{R}^n$. This idea was used as a basis to study the dynamics of the immune response, with $n$ assumed to be a small number \cite{segel1989shape,de1992pattern}. Subsequent work to estimate $n$ based on the frequency of cross reactivity resulted in the conclusion that $n$ was around five to eight \cite{smith1997deriving}.

The notion of antigenic space has proven particularly popular for understanding the evolution of influenza H3N2 \cite{fonville2014antibody}. This strain has been circulating in the human population since 1968 and gradually mutating. These mutations can in principle be represented as the movement of the virus through antigenic space. As the antigen moves
it can evade the antibodies elicited by older strains and thus reinfect individuals.
The distance between a viral strain and an antibody can be measured via the hemagglutination inhibition (HAI) assay, in which a viral strain and a serum of antibodies are both added to a culture of red blood cells. If the
antibodies are ineffective against the viral strain then the virions stick to the red blood cells causing them to cluster together (hemagglutinate). However if the antibodies are effective, they will neutralize the virions and inhibit their hemagglutination of the red blood cells. In the former case, the strain and the serum are distant antigenically, whereas in the latter case they are close. By performing serial dilutions of the antibody serum, one can quantify just how close a serum and antibody are.

Points in antigenic space can be inferred from a distance matrix via multidimensional scaling (MDS). 
Low dimensional reconstructions of antigenic space can reproduce the HAI data with high fidelity, and adding new dimensions beyond $n=5$ does not improve
the quality of the fit \cite{lapedes2001geometry,smith2004mapping}. Thus it may be tempting to conclude that influenza is evolving in an antigenic space of no more than five dimensions.

\subsection{Outline of results}
In this work we will argue that influenza H3N2 is evolving in a very high dimensional space, and that it may appear to be low dimensional due to the nature of random walks. Our argument consists of three parts.
\begin{enumerate}
\item High dimensional random walks contain most of their variance in a small number of dimensions. Specifically, one would expect at least 60\% of the variance
to occur in a single dimension. We show this via principal component analysis. However, we note that the true dimensionality of random walks can be revealed by out of sample PCA.
\item This apparent low dimensionality also occurs with multidimensional scaling --- the method used to analyze HAI titer data. We simulate HAI data generated using a high dimensional random walk and find we can accurately reproduce the data with points taken from a low dimensional space; increasing the dimensionality of our representation
does not improve our fit. We also show that non-metric MDS can be used to reconstruct an infinite dimensional random walk in a single dimension.
\item Finally, we show that H3N2 data has characteristics of a high dimensional random walk, indicating that it was unlikely to result from a random walk of dimensionality less than $n=10$. This is even the case when we consider that the random walk of H3N2 is likely self avoiding.
\end{enumerate}

\subsection{Why a random walk?}
Throughout this paper we argue for a high dimensional random walk as a model for influenza evolution. A random walk may seem \textit{a priori} to
be a poor model for viral evolution, as immunological memory should prevent a virus from revisiting areas of antigenic space. Therefore we should expect the path of viral evolution to be self avoiding. In high dimensions an unbiased random walk and self avoiding random walk will behave very similarly, because a high dimensional random walk is already extremely unlikely to cross itself.
In an $n$ dimensional random walk the distance between points $i$ and $j$ is a random variable $D_{ij}$. Its distribution is
\begin{align*}
D_{ij}^2\sim\alpha |i-j|\chi^2_n
\end{align*}
where $\alpha$ is a constant of proportionality. This means that for large $n$ the distances increase in a very predictable manner as the $\chi_n^2$ distribution narrows. The probability of the random walk approaching a previous point is essentially zero, so we need not include any further tendency for self avoidance. However, in the latter part of the paper we will
address the question as to whether low dimensional self avoiding random walk could also be consistent with the data.

\subsection{True dimensionality vs effective dimensionality}
Let $x_i \in \mathbb{R}^n$ represent distinct viral strains and/or antisera.
The antigenic dissimilarity of the two different strains $x_i$ and $x_j$
is the euclidean distance $D_{ij}=\|x_i-x_j\|_2$. $D_{ij}$ represents the true distance between
these two strains and $n$ is the true dimensionality of antigen space.

If we can represent each point $x_i$ with a corresponding point $y_i \in\mathbb{R}^k$
such that the distances from $y_i$ to $y_j$ is approximately $D_{ij}$, then we say that the effective dimensionality of $\{x_i\}$ is $k$.

There are several possible ways to obtain the points $y_i$ and to evaluate how closely the reconstructed distances match the true or measured distances. In this paper we shall make use of three: classical MDS, metric MDS and nonmetric MDS. Classical MDS is also commonly known as principal component analysis (PCA). The points $y_i$ are found by orthogonally
projecting the points $x_i$ onto the dimensions that contain the most variance.
Metric MDS simply minimizes the residual between the true distances and the reconstructed distances. Non-metric MDS is similar to metric MDS but the relationship
between the true distances and the reconstructed distances is only assumed to be monotonically increasing as opposed to directly proportional.

\section{Results}
\subsection{Truly high dimensional random walks have low effective dimensionality according to PCA}
We will now show that random walks with high true dimension can have a very low effective dimension. Our analysis will focus
on infinite dimensional random walks, but we will show via simulations that finite random walks have similar behavior.
Let $x_i\in\mathbb{R}^n$ be the $i$th step in our random walk in $n$ dimensions.
Let $X$ be an $m$ by $n$ matrix whose $i$th row is $x_i$. The rows of $x_i$ are determined by a seres of random steps $b_i \in \mathbb{R}^n$,
each entry of which is an independent sample from $\mathcal{N}(0,1/\sqrt{n})$
\begin{align}
x_1&=b_1\\
x_{i}-x_{i-1}&=b_i \quad i>1
\end{align}
We can state this relationship compactly as $LX=B$, where the rows of $B$ are $b_i$ and $L$ is an $m$ by $m$ matrix with one on the diagonal
and negative one on the subdiagonal.

To perform PCA, we first center each column of $X$ by premultiplying be the projection matrix $P=I-\one^T\one$. Note that centering each column is just a translation of the data, so all pairwise distances
are preserved. We then compute the eigenvalues of $A=PXX^{T}P$, $\lambda_1>\lambda_2>\cdots \lambda_m$. These eigenvalues indicate the variance in each principal
component of $X$. For this section we shall define effective dimension as
\begin{align}
\quad \text{ED}_a=\min \left\{d:\sum_{i=1}^k \lambda_k\geq a~ \text{tr}(A)\right\}
\label{qeqn}
\end{align}
where $0<a<1$ is the fraction of variance that we require within the first $k$ dimensions.

The random matrix $A=PL^{-1}BB^{T}L^{-T}P$ is full rank and has a wishart distribution if the number of steps is no greater than the number of dimensions, i.e. $m \leq n$. If number of steps is greater than
the number of dimension ($n>m$), $A$ is singular and has a pseudo-wishart distribution.

\subsubsection{Infinite dimensional random walks}
The matrix $W=BB^T$ is an uncorrelated Wishart matrix. The joint probability distribution of the eigenvalues of Wishart matrices is known, but the formula is cumbersome \cite{johnstone2001distribution, kang2003largest, zanella2009marginal}. Therefore we will focus our analysis on the limiting behavior as $n\rightarrow \infty$, i.e. infinite dimensional random walks.
 As $n$ increases, $W$ approaches the $m$ by $m$ identity matrix. (To see this, either note that the $i,j$th entry of $W$ is $<b_i,b_j>$). Using this simplification, the matrix $A$ approaches
\begin{align}
A=PL^{-1}L^{-T}P
\end{align}
Noting that $L^{-1}$ is an upper triangular matrix with all ones, we can calculate $tr(A)=(m-1)(m+1)/6$.

We can first observe that this matrix has one eigenvalue $v=\one$ with corresponding eigenvalue $\lambda_1=0$. To find the remaining eigenvalues we use
the pseudo-inverse of $A$, $\pseudo{A}$ which has the simple form
\begin{align}
\pseudo{A}=
\begin{pmatrix}
1 & -1 & 0 &\cdots & 0\\
-1 & 2 & -1 & \ddots &\vdots\\
0 & \ddots & \ddots & \ddots & 0\\
\vdots & \ddots & -1 &2 & -1\\
0 & \cdots &0& -1 & 1
\end{pmatrix}
\end{align}
The eigenvalues of this matrix are $\mu_s=2-2\cos(\pi s/m)$. Therefore the eigenvalues of $A$ are
\begin{align}
\lambda_k=1/(2-2\cos(\pi s/m))
\end{align}

\begin{figure}
\includegraphics[width=\textwidth]{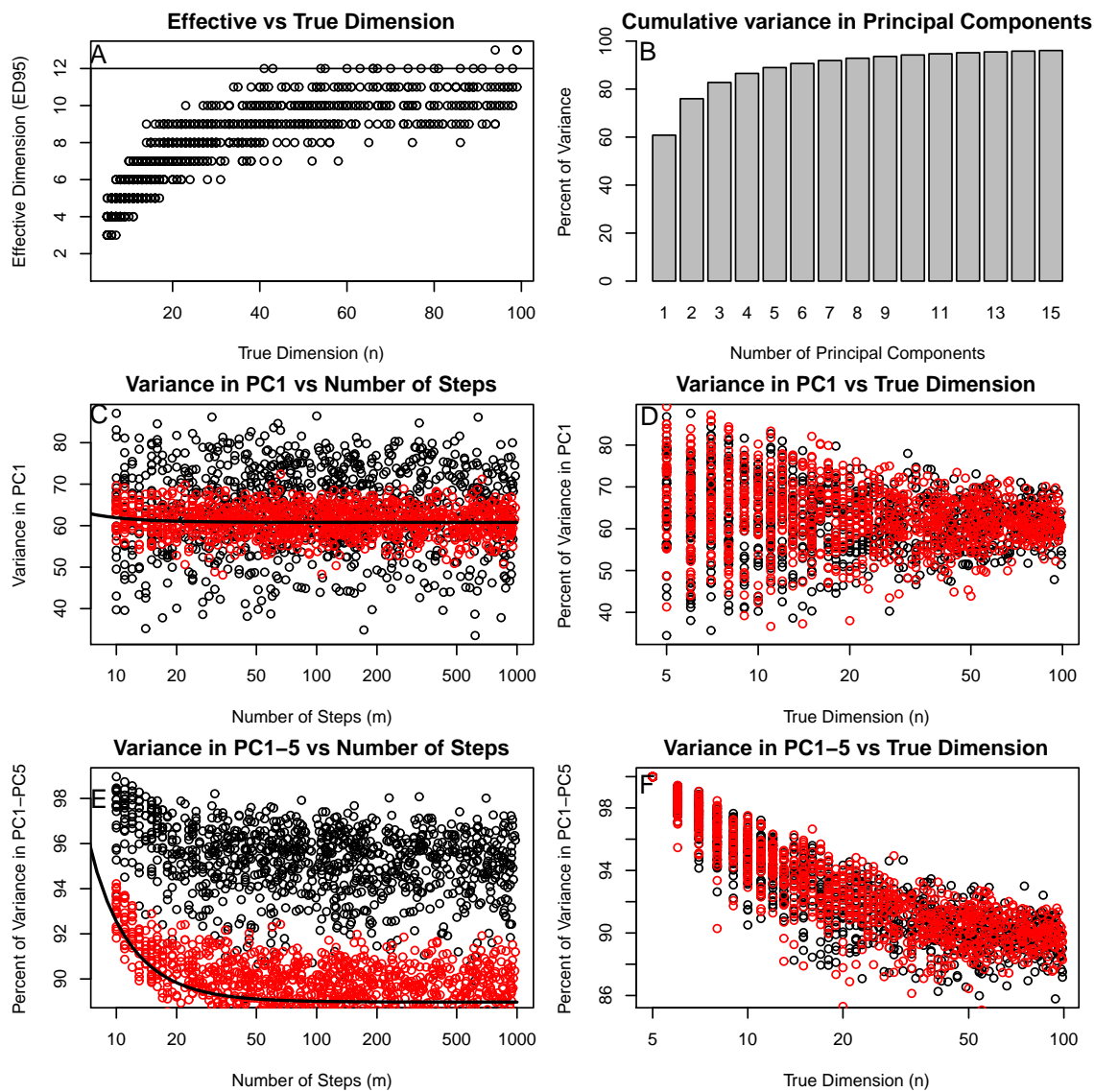}
\label{fig1}
\caption{\textbf{Behavior of random walks} A: The effective dimensionality (the number of dimensions containing 95\% of the variance) is always low, even for high dimensional
random walks. The horizontal line shows the theoretical limit of the effective dimensionality as $n\rightarrow \infty$.
B: The cumulative percentage of the total variance in the first 15 principal components or an infinite dimensional random walk.
C/E: Percentage of variance in the first (C) or first five (E) principal components vs number of steps in 10-dimensional (black) or 100-dimensional (red) random walks.
D/F:Percentage of variance in the first (D) or first five (F) principal components vs number of dimension in 100-step (black) or 1000-step (red) random walks.}
\end{figure}

The fraction of the variance in the first $k$ dimensions is
\begin{align}
\frac{\sum_{s=1}^k \lambda_s}{\text{tr}(A)}&=\frac{\sum_{s=1}^k (1-\cos(\pi s/m))^{-1}}{(m-1)(m+1)} &\text{Finite $m$}\\
&=\frac{6}{\pi^2}\sum_{s=1}^k 1/s^2 &\text{Infinite $m$}
\end{align}
For infinite dimensional random walks, the first principal component contains at least  $6/\pi^2$ or roughly 60\% of the total variance. The first two components contain 80\% of the variance, but the subsequent convergence is slow as 12 components are needed to account for 95\% of the variance (Fig. \ref{fig1}A,B). These numbers are the limiting behavior as the length of the random walk grows. For shorter walks, more of the variance is in the first few components (Fig. \ref{fig1}C, E) and thus the effective dimensionality may be even lower.

\subsubsection{Finite dimensional random walks}
In a finite number of dimensions $n$, the matrix $A$ is a random variable with a correlated Wishart or Pseudo-Wishart distribution. We therefore use simulation to investigate the effective dimensionality. As the true dimensionality increases, the effective dimensionality asymptotically approaches the value for the infinite case, in which 12 dimensions contain 95\% of the variance. However, the effective dimensionality for finite random walks is stochastic and in a few cases we find that it may exceed 12, although the mean is lower (Fig \ref{fig1}A).

Like the effective dimensionality, the percentage of variance in the first principal component is a random variable, $\Psi_1$. Regardless of the length of the walk or the true dimensionality, the mean of $\Psi_1$ is very similar to the infinite dimensional case, and the variance of $\Psi_1$ decreases with the true dimensionality but is independent on the length of the walk (Fig \ref{fig1}C,D). This gives the somewhat counter intuitive result that lower dimensional
random walks may have less variance in their first principal component than higher dimensional walks.
When we consider the percentage of variance in the first 5 components $\Psi_5$, this pattern disappears (Fig. \ref{fig1}E,F). The values of $\Psi_5$ tend to decrease monotonically
as both the true dimensionality and the number of steps increase, asymptotically approaching the prediction of roughly 89\% from the infinite dimensional case.

\subsubsection{True dimensionality of random walks can be revealed by subsampling}
We have demonstrated that when we perform PCA on a random walk, we find most of the variance in only one dimensoin.
Therefore, it is be difficult to tell whether the data originate from a high or low dimensional random walk using PCA. In this section we demonstrate a way around this problem
using out of sample PCA.

\begin{figure}
\includegraphics[width=\textwidth]{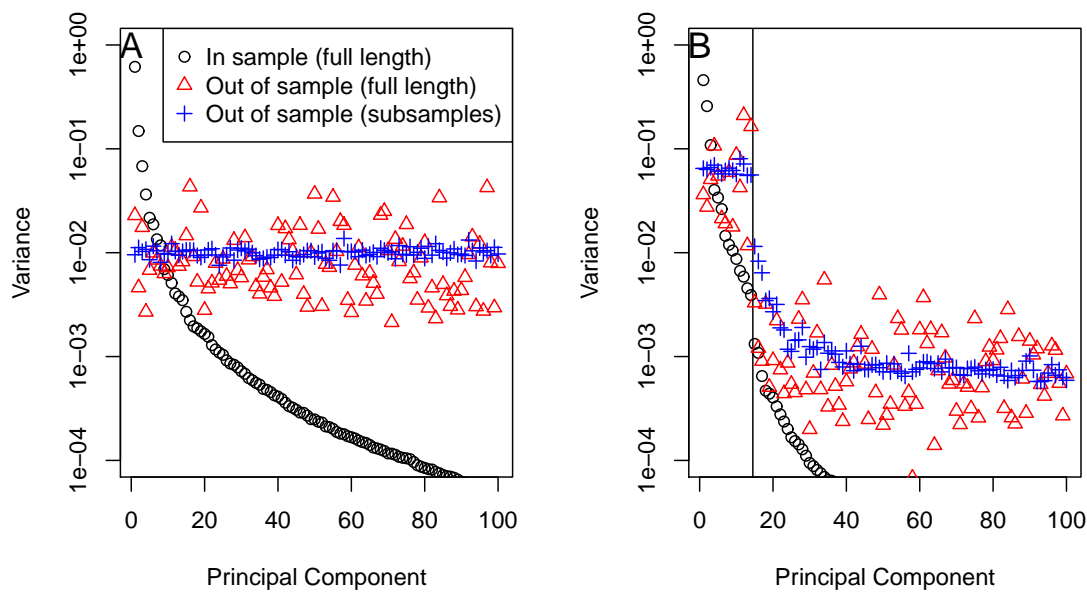}
\label{fig5}
\caption{\textbf{Out of sample PCA can recover true dimensionality} We can recover the true dimensionality of a random walk by performing out of sample PCA. We split the walk into two halves. Performing PCA on the first half tends to overestimate the variance in the first few principal components (black dots). We then use the projection matrix derived fromPCA on the first half to both the full second half of the random walk (red triangles) or length 25 subsequences from the second half (blue crosses). We show the variance estimated for each principal component in A) a uniform random walk in 100 dimensions and B) a random walk 100 dimension where the first 14 dimensions have 100 times as much variance as
the remaining 86.}
\end{figure}

We consider two scenarios: a uniform random walk in which movement in every direction is equally likely, and a nonuniform random walk in which movement is ten times greater --- and thus the variance is 100 times greater --- in 14 dimensions compared to the other 86. PCA finds the majority of variance is in the first dimension for both of these scenarios (Fig~\ref{fig5}).
However if we take the projection matrix derived from performing PCA on the first half of the random walk and apply it to the second half, we see that the bias towards the first dimension largely disappears. The approximation improves further if we divide the second half into subsequences, apply the projection to each subsequence and then average the results (see methods for further details). This latter method shows clearly that in the uniform random walk, each principal component has roughly the same variance. It also captures the fact that
14 dimensions contain significantly more variance in the nonuniform random walk. Therefore out of sample PCA can in principle allow one to recover the true dimensionality of a random walk.

\subsection{Truly high dimensional random walks look low dimensional according to MDS}
\subsubsection{H3N2 evolution}
Derek Smith, Alan Lapedes and colleagues have used dimensional reduction techniques to reconstruct two-dimensional antigenic maps of the evolution of Influenza H3N2 \cite{smith2004mapping}.
In these maps, some points represent viruses whereas others represent antisera. Distances can only be computed between a virus and an antisera as opposed to between different
viruses directly. Thus classic MDS, which requires measured distances between all pairs of points, cannot be used for this problem. Instead, they use metric MDS with a slight modification to handle
sub-threshold values. Using this method, they have reconstructed a two-dimensional map of antigenic space. To validate this map, they challenged it to predict
the antigenic distance between strains and antisera that were not used as inputs to build the map. They report a good correlation (roughly 80\%) between the predicted and actual
values, and furthermore that the accuracy of their predictions is similar for a two dimensional map and for higher dimensions. This is consistent with prior work showing that no more than five dimensions is required to satisfactorily reconstruct HAI data \cite{lapedes2001geometry}. 

Although these authors make no claims
about the true dimensionality of the underlying space, it may be tempting to conclude that antigenic space has a relatively low dimensionality.
Our work, however, suggests that the antigenic space of H3N2 may appear to have low dimensionality even though the true dimensionality is very high. We therefore created artificial data to to mimic the evolution of H3N2 via a high dimensional random walk.  We then followed the procedure for evaluating the quality of reconstructions outlined in \cite{smith2004mapping}.

\begin{enumerate}
\item We partitioned the real and artificial distance matrices into a training set (90\% of the data) and a test set (10\%)
\item Using only the distances in the training set we reconstructed points in antigenic space corresponding to each virus and serum. We varied the dimensionality of antigenic space ($k$) between 2 and 20. Note our artificial data was created using points in 100 dimensional space.
\item We then uses the reconstructed points to predict the antigenic distances in the test set. We evaluated the quality of these fits both by the root-mean-square error and the correlation
between the actual and predicted distance. These were the same criteria used in \cite{smith2004mapping}.
\end{enumerate}

\begin{figure}
\includegraphics[width=\textwidth]{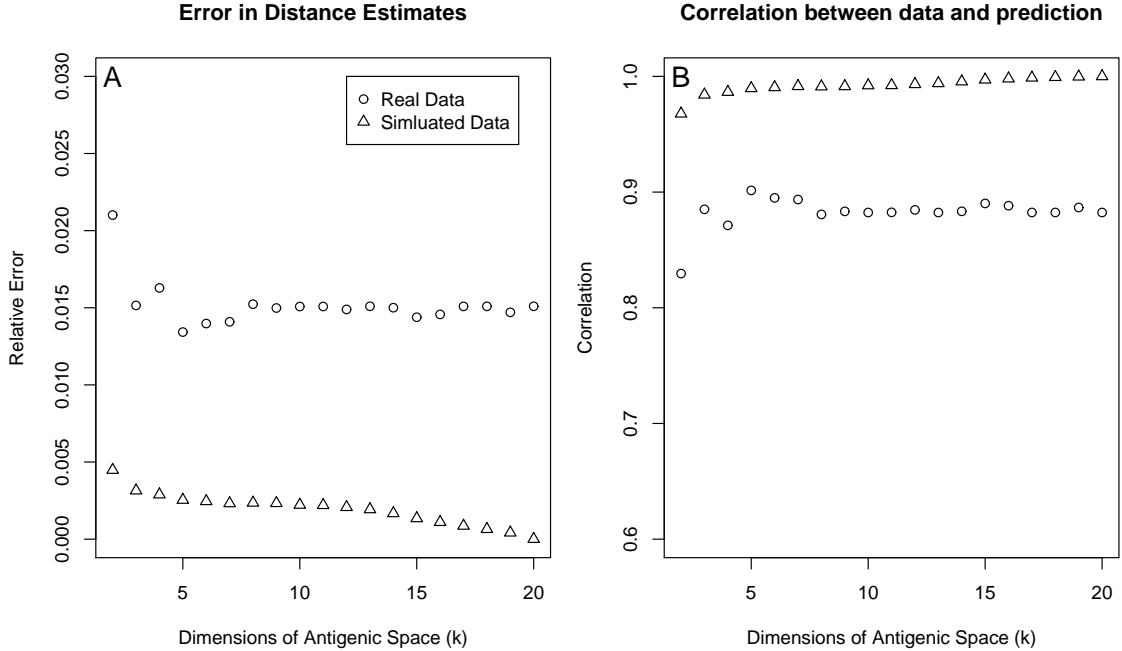}
\label{fig2}
\caption{\textbf{High dimensional walks can look low-dimension} We used multidimensional scaling to reconstruct maps of up to 20 dimensions from a supplied distance matrix.
We generated maps from H3N2 HI titer data, as well as simulated data generated from a 100 dimensional random walk. We tested the accuracy of these maps by training them on 90\% of the data and then using them to predict the remaining 10\%. We plot the relative error (A) and correlation between data and prediction (B) versus the number of dimensionality of the map.}
\end{figure}

We find that increasing $k$ beyond 2, i.e. adding more dimensionality to our reconstructed antigen space, did not improve the quality of the distance estimates (Fig \ref{fig2}A,B). This suggests that even data originating from a high dimensional space can be approximated in two dimensions.

\subsubsection{An extreme case: High dimensional random walks can look one dimensional in non-metric MDS}
The goal of metric MDS is to find points $x \in \mathbb{R}^k$ whose pairwise euclidean distances approximately match a set of given distances. In particular, the points $x$ must minimize the stress
\begin{align}
\text{Metric Stress}=\sqrt{\frac{(\|x_i-x_j\|-d_{ij})^2}{\sum d^2_{ij}}}
\end{align}
where $d_{ij}$ is the distance between point $i$ and $j$.

In practice, the $d_{ij}$ used as an input to MDS may not represent physical distances but rather a generic measure of dissimilarity that does not behave like a metric. For example,
consider three points with dissimilarities $d_{12}=1$, $d_{23}=2$ and $d_{13}=100$. The dissimilarity cannot be a metric as it violates the triangle inequality. However, there may be an
underlying metric space that generated the points, and the underlying dissimilarity represents a transformation, $f$, of that metric.
The goal of non-metric MDS is to find both $f$ and values for $x$ that minimize
\begin{align}
\text{Non-metric Stress}=\sqrt{\frac{(f(\|x_i-x_j\|)-d_{ij})^2}{\sum d^2_{ij}}}
\label{nonmet}
\end{align}

The distance-squared between steps in an $n$ dimensional random walk follows a $\chi^2_n$ distribution.
\begin{align*}
D_{ij}^2\sim\alpha |i-j|\chi^2_n
\end{align*}
where $\alpha$ is a constant of proportionality. In this case we let the dissimilarity $d_{ij}$ be the euclidean distance
$D_{ij}$.
As $n \rightarrow \infty$, the dissimilarity becomes exactly proportional to $\sqrt{|i-j|}$. Thus we can construct a one dimensional map of our random walk
via
\begin{align}
\begin{split}
\text{Dissimilarity between point $i$ and point $j$} \,\,\quad d_{ij}^2 &=\alpha |i-j|\\
\text{Location of points} \qquad x_i &=\alpha i\\
\text{Monotonic Transform} \quad f(\cdot) &=\sqrt{\cdot}
\end{split}
\label{interpointd}
\end{align}
Plugging \eqref{interpointd} into \eqref{nonmet} we find that the stress is identically equal to zero. Therefore an infinite dimensional random walk can
be exactly represented in one dimension using non-metric MDS.

\subsection{Distinguishing between low dimensional and high dimensional random walks using distance measures}
\begin{figure}
\includegraphics[width=\textwidth]{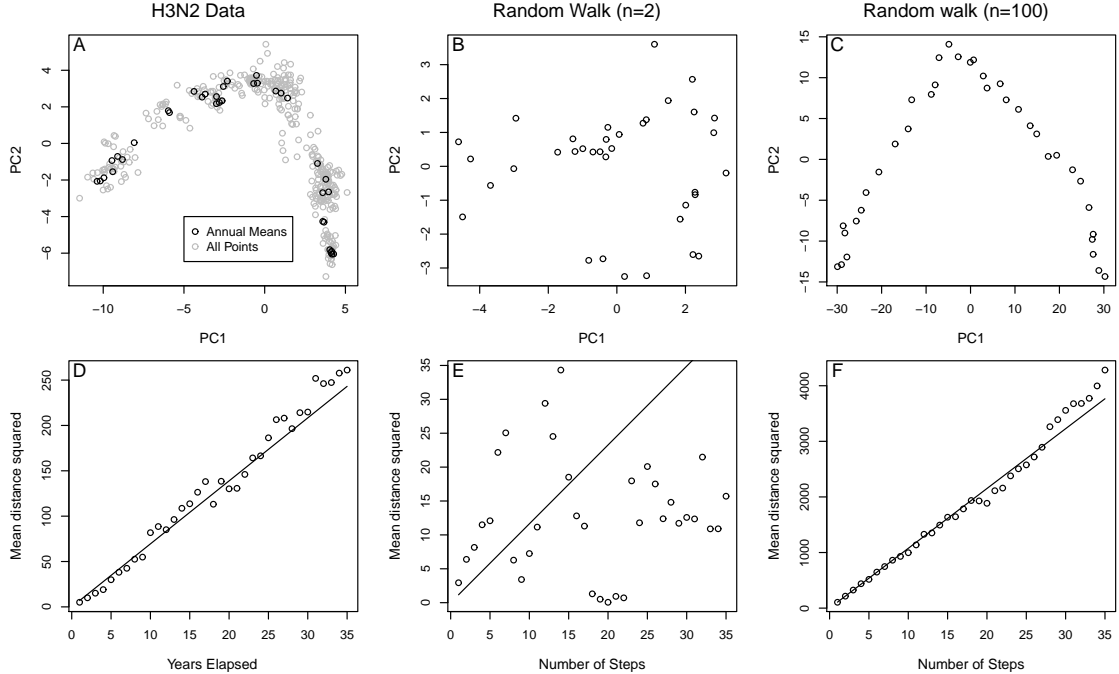}
\label{fig3}
\caption{\textbf{Distinguishing between high and low dimensional random walks} High dimensional random walks follow a characteristic parabolic curve when
plotting the first two principal components (A,B,C). Furthermore, the relationship between number of steps and distance squared between two points is linear (D,E,F).
The H3N2 antigen map (A,D) resembles a 100-dimensional random walk (C,F) more than a 2-dimensional random walk (B,E).}
\end{figure}

So far we have demonstrated that high dimensional random walks can appear low dimensional using methods such as MDS or PCA. We therefore ask whether there is any
good way to distinguish a low dimensional from a high dimensional random walk. One characteristic of high dimensional random walks is the curve that forms when the first two principal components are plotted against each other \cite{bookstein2012random}. This pattern emerges as the number of dimensions approaches infinity. Fig. \ref{fig3}A shows the first two principal
components by year, whose path is reminiscent of the quadratic shape predicted in \cite{bookstein2012random}. This pattern is suggestive of a high (FIg \ref{fig3}B), rather than low dimensional (Fig \ref{fig3}C) random walk.

We next sought an objective measure to distinguish a low dimensional from high dimensional walk.
Once again we take advantage of the fact that the square of the distance between the $i$th and $j$th step in an $n$ dimensional random walk is proportional to a $\chi^2$ distribution with $n$ degrees of freedom.
\begin{align}
D_{ij}^2\sim\alpha |i-j|\chi^2_n
\end{align}
Let $\bar{D}^2_s$ denote the average distance between two points separated by $s$ steps, i.e.
\begin{align}
\bar{D}^2_s=\frac{1}{m-s}\sum_{i=1}^{i-m+s}D^2_{i, i+s}
\end{align}
where $m$ is the number of steps.
As $n\rightarrow \infty$, the coefficient of variation of $\chi^2_n \rightarrow 0$, so we should expect $\bar{D}^2_s \approx \alpha s$ for some $\alpha$.
In other words, for high dimensional random walks plotting $\bar{D}^2_s$ vs $s$ should give a linear relationship.
We therefore grouped all antigenic map points in the H3N2 data by year, representing both viral strains and anti-sera, and computed the mean of each group.
Computing the distance-squared between the annual means and plotting versus the number of years elapsed does indeed produce a linear pattern. In this regard, the H3N2 data once again qualitatively resembles a high dimensional random walk rather than a low dimensional one.(Fig \ref{fig3}D-F).

\begin{figure}
 \includegraphics[width=\textwidth]{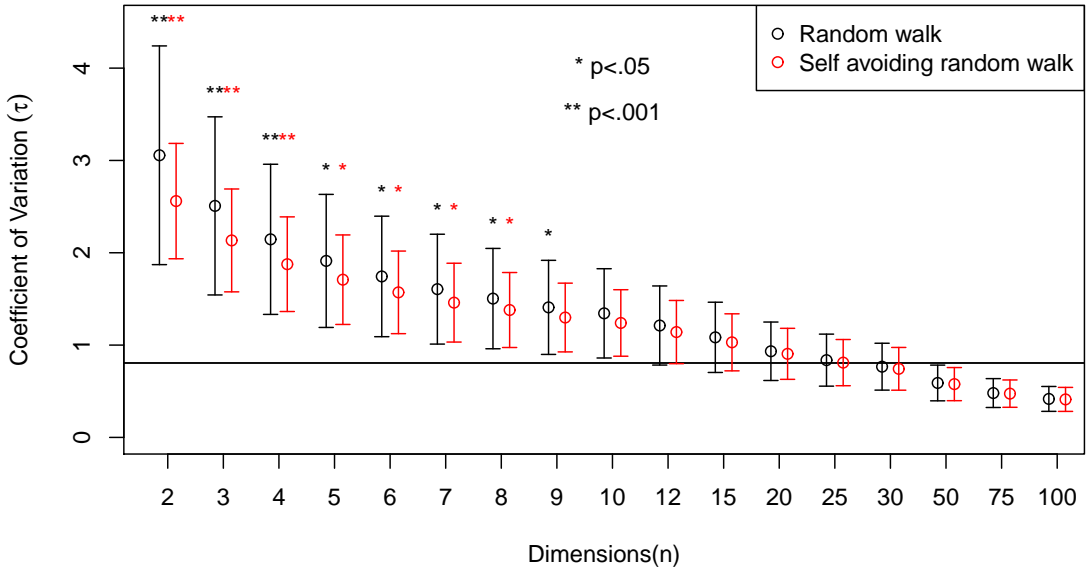}
\label{fig4}
\caption{\textbf{True dimensionality of antigenic is at least $n=10$} We measured deviation from a linear relationship between number of steps and distance squared between points by dividing the latter by the former and taking the coefficient of variation. The horizontal line shows the value found for the H3N2 antigenic map, which is compared
to the values found for 10000 simulated random walks of each of the indicated dimensions. We find that the consistency of the linear trend as depicted in \ref{fig3}D is inconsistent
with a random walk of dimension less than 10.}
\end{figure}

To quantify this resemblance, let $\tau$ be the coefficient of variation of $x_s=\frac{1}{s}\bar{D}^2_s$, i.e
\begin{align}
\tau=\sqrt{\frac{\sum_{s=1}^{m-1} x^2_s}{\left(\sum_{s=1}^{m-1} x_s\right)^2}-1}
\end{align}
 For high dimensional random walks,
we expect $\tau$ to be close to zero. We simulated $\tau$ for large numbers of random walks and compared to the value computed for the H3N2 data (Fig \ref{fig4}).
We find that the H3N2 data is unlikely to have been produced by a random walk of less than ten dimensions ($n<10$). Note also that the value of $\tau$ for the data may be inflated due to any number of sources of variability not accounted for in the model, such as measurement error or a heavy tailed distribution in annual step size. Therefore $d=10$ is the lower bound for the dimensionality antigenic space and dimensionality of 30 or more is quite likely. These findings extend to self avoiding random walks, as decribed in the methods section, which behave similarly in this regard.

\section{Discussion}
High dimensional random walks can appear to be low dimensional. We have shown that most of their variance will appear in a single principle component, and that their pointwise distances can be well approximated by embedding in a two dimensional space. Therefore when using principal component analysis (PCA) or multidimensional scaling (MDS) we may erroneously find that only a few dimensions are important. We demonstrate that by splitting the data set into two, we can use out of sample PCA to potentially recover the true dimensionality of the random walk. If only MDS is possible, perhaps due to missing data, we demonstrate that high dimensional and low dimensional random walks can be distinguished by testing the relationship between distance and step number.

We apply these ideas to the evolution of influenza H3N2. Using data from the HAI assay different strains from 1968 to 2003 can be readily represented in a two dimensional
antigenic map. The distances on this map represent the level of cross immunity that an antibody response to one strain may provide against another. Given this it may be tempting to conclude that influenza is constrained to move in a two dimensional space or at least is biased to move in two dimensions. However, we caution that antigenic space may in fact be very high dimensional or even infinite dimensional and we would still expect to be able to reconstruct it in two dimensions. In fact, we find that a low dimensional random walk or self avoiding random walk is very unlikely to have produced the H3N2 HAI titer data.

One interesting aspect of influenza H3N2 evolution is that despite the annual variation in strain, the strains have not been diverging. Different strains circulating in a given year tend to be close together antigenically. There have been many models proposed to explain this phenomenon \cite{gog2002dynamics,ferguson2003ecological,koelle2006epochal,bedford2012canalization,wikramaratna2013antigenic}. Here we have only shown that the data is consistent with a high dimensional random walk or self avoiding random walk. We assume that the direction of the next step is equally likely to be in any direction, provided that this does not bring the walk back to an already visited region. Therefore our model is consistent with antigenic drift hypotheses as described in \cite{gog2002dynamics,ferguson2003ecological,koelle2006epochal,bedford2012canalization} but not the antigenic thrift hypothesis described in \cite{wikramaratna2013antigenic}.

\section{Methods}
\subsection{Out of sample PCA}
\begin{enumerate}
\item Start with a length $2m$ random walk in $n$ dimensions.
\item Split the random walk into two halves.
\item Apply PCA to the first half. This will yield an orthogonal projection matrix $P\in\mathbb{R}^{n\times n}$ which projects any point onto the principal component axis.
\item Split the second half of the walk into subsequences with a short length, $\mu$. These subsequences consist of steps $m+1$ to $m+\mu$, $m+\mu+1$ to $m+2\mu$, etc. Let $S_i \in\mathbb{R}^{n\times \mu}$ denote the $i$th such subsequence.
\item Apply $P$ to each subsequence $S_i$, i.e. compute $B_i=PS_i$.
\item Let $\mathbf{\sigma^2}_i$ be the variance of each row of $B_i$.
\item Let $\mathbf{\sigma^2}=\frac{1}{\nu}\sum_{i=1}^{\nu}\mathbf{\sigma^2}_i$ be the sum of all $i$, where $\nu$ is the number of subsequences (i.e. $\nu\mu=m$).
\end{enumerate}

\subsection{Generating simulated data}
We want to generate $N_v$ viral strains and $N_s$ antisera. We will represent each strain and antisera as a point in $\mathbb{R}^n$, where $d$ is our chosen dimensionality of antigenic space. The generation of the artificial data makes repeated use of random vectors in $\mathbb{R}^n$. We draw these random vectors from $N(0,I_d)$, i.e. their
entries are i.i.d unit normal random variables.
We first generate the viral strains via a random walk. Let $V_i \in \mathbb{R}^n$ represent the $i$th viral strain in the walk, then
\begin{align*}
V_1&=b_1\\
V_{i}-V_{i-1}&=b_i \quad i>1
\end{align*}
where $b_i~N(0,I_n)$

Next we use these strains to randomly generate antisera. Let $S_j \in \mathbb{R}^n$ be the $j$th antiserum, then
\begin{align*}
S_j&=V_{p_j}+b
\end{align*}
where $p_j$ is a random integer between 1 and $N_v$ and $c_j~N(0,I_n)$.

\subsection{Multi Dimensional Procedure}
In this paper we use the form of multidimensional scaling described in \cite{smith2004mapping}. We perform this scaling in two distinct cases: starting from
a matrix of measured HI and starting from artificial data generated as described in the previous section.

Not every viral strain is measured against each serum. Let $N_{\text{Meas}}$ be the number of measurements and let $0<l\leq N_{\text{Meas}}$ index those measurements.
We can then define
\begin{align*}
i_l & \quad \text{Strain corresponding to the $l$th measurement}\\
j_l & \quad \text{Serum corresponding to the $l$th measurement}\\
d_l & \quad \text{Distance between strain $i_l$ and serum $j_l$}
\end{align*}

In the case of real HI titer data, we convert the titer to a distance using the method described in \cite{smith2004mapping}.
When using artificial data, we simply compute $d_l=\|V_{i_l}-S_{j_l}\|$.

Next we seek a set of points $\hat{V}_i\in \mathbb{R}^k$ and $\hat{S}_j\in \mathbb{R}^k$ that minimizes the objective function
\begin{align*}
F\left(\{\hat{V}_i\},\{\hat{S}_j\}|\{d_{l}\}, \{i_l\}, \{j_l\}\right)=\sum_{l=0}^{N_{\text{Meas}}} \left(d_l-\|\hat{V}_{i_l}-\hat{S}_{j_l}\|\right)^2
\end{align*}
We use values of $k$ between 2 and 20.

We minimize $F$ numerically using the optim function in R. We use the BFGS method and provide the analytically derived gradient of $F$ for efficient computation.
This is an iterative algorithm that requires an initial guess. We generate these guesses using classic MDS, which is equivalent to principal component analysis. The classic MDS algorithm requires a complete distance matrix. In the case of the artificial data, we can calculate this distance matrix directly. In the case of the real data, we construct a plausible distance matrix by finding the shortest path between any two points whose distance was not directly measured. Note that although this plausible distance matrix is likely incorrect, it is only used to find a good starting guess for our algorithm. The solution found using this method compared favorably to the solution found in \cite{smith2004mapping}

\subsection{Self avoiding random walk}
The self avoiding random walk behaves similarly to an unbiased random walk accept that the new steps are excluded from getting to close to previous ones.
Each new step in the walk is determined according to the following algorithm.
\begin{enumerate}
\item Given a current step $x_i$, propose a new step $\hat{x}=x_i+b\in \mathbb{R}^n$, where the entries of $b$ are i.i.d with distribution $\mathcal{N}(0,1/\sqrt{n}$.
\item Calculate the probability of acceptance via
\begin{align*}
P&=\prod_{j=0}^i g(\hat{x}|x_i)\\
g(\hat{x}|x_i)&=1-e^{-\|\hat{x}-x_i\|^2}
\end{align*}
\item With probability $P$ accept the new step and let $x_{i+1}=\hat{x}$. With probability $1-P$ return to step one and propose a new step.
\end{enumerate}

\bibliographystyle{alpha}
\bibliography{PCA}
\end{document}